\documentclass{INTERSPEECH2023}

\usepackage{amsmath}
\usepackage{graphicx}
\usepackage{arydshln}
\usepackage{xcolor}
\usepackage{multirow}
\usepackage{amsfonts}
\usepackage{booktabs}
\usepackage[vlined,linesnumbered,ruled]{algorithm2e}

\newcommand{\ra}[1]{\renewcommand{\arraystretch}{#1}}

\interspeechcameraready

\title{Factual Consistency Oriented Speech Recognition}
\name{Naoyuki Kanda, Takuya Yoshioka, Yang Liu}
\address{
  Microsoft, USA}
  
\email{\{nakanda,tayoshio,yaliu10\}@microsoft.com}

\begin{document}

\maketitle
 
\begin{abstract}
%
This paper presents a novel optimization framework for automatic speech recognition (ASR) with the aim of reducing hallucinations produced by an ASR model. The proposed framework optimizes the ASR model to maximize an expected factual consistency score between ASR hypotheses and ground-truth transcriptions, where the factual consistency score is computed by a separately trained estimator. Experimental results using the AMI meeting corpus and the VoxPopuli corpus show that the ASR model trained with the proposed framework generates ASR hypotheses that have significantly higher consistency scores with ground-truth transcriptions while maintaining the word error rates close to those of cross entropy-trained ASR models. Furthermore, it is shown that training the ASR models with the proposed framework improves the speech summarization quality as measured by the factual consistency of meeting conversation summaries generated by a large language model. 
\end{abstract}

\section{Introduction}

Thanks to substantial progress in deep learning techniques, such as advanced model architectures \cite{vaswani2017attention,gulati2020conformer} and training criteria \cite{graves2014towards,povey2016purely,shannon2017optimizing,prabhavalkar2018minimum} as well as the utilization of large-scale training data \cite{parthasarathi2019lessons,kanda2021large,zhang2022bigssl,radford2022robust}, the automatic speech recognition (ASR) accuracy has significantly improved. For example, latest ASR systems achieved lower word error rates (WERs) than human transcribers for many public test sets, such as the LibriSpeech \cite{panayotov2015librispeech,amodei2016deep} and Switchboard benchmarks \cite{godfrey1992switchboard,xiong2016achieving}. It was also reported that a large ASR model trained on 68K hours of transcribed data curated from the Web achieved domain robustness and accuracy closer to that of the human transcribers \cite{radford2022robust}.

However, the ASR technology is still prone to recognition errors under different conditions. 
To complicate matters, an ASR model with a strong language modeling (LM) capability
can produce fluent but factually ungrounded, or hallucinated, ASR errors, 
regardless of whether the LM is part of an end-to-end ASR model \cite{graves2006connectionist,graves2012sequence,chorowski2014end} or explicitly provided 
as with hybrid ASR \cite{seide2011conversational}. 
These errors often make the transcriptions difficult or confusing for humans and downstream natural language processing (NLP) modules, such as an automatic summarization module,  to comprehend and thus have a detrimental effect. 

Many approaches were explored to alleviate the negative effect that the ASR errors have on the downstream tasks. A conventional approach is using N-best hypotheses or lattices in the subsequent NLP modules \cite{hakkani2006beyond,morbini2012reranking}. However, feeding the N-best/lattice representations requires significant changes to the NLP modules, and how to do so is not obvious for many state-of-the-art NLP models pre-trained on regular texts, such as GPT-3 \cite{brown2020language}.
Another approach is joint optimization of the ASR and NLP modules either by end-to-end modeling \cite{haghani2018audio,rao2020speech,sharma2022end,shon2022slue} or by using an NLP task-oriented loss \cite{rao2021mean}. 
While this approach has shown promising improvements on the downstream NLP tasks,
 using a task-specific architecture or loss may make the ASR model suitable only for the specific task it was trained for. 
 In addition, joint modeling approach is difficult to apply to a task that handles very long sequences, such as a speech summarization task. For example, the current text summarization model is often trained to process more than 8,000 tokens, which roughly corresponds to 30--40 audio minutes. It is not trivial to train joint models for such long audio inputs, and therefore existing joint ASR and  summarization models are limited to handling short durations such as 100 seconds \cite{sharma2022end} or 300 seconds \cite{shon2022slue}.

\begin{figure}[t]
  \centering
  \includegraphics[width=1.0\linewidth]{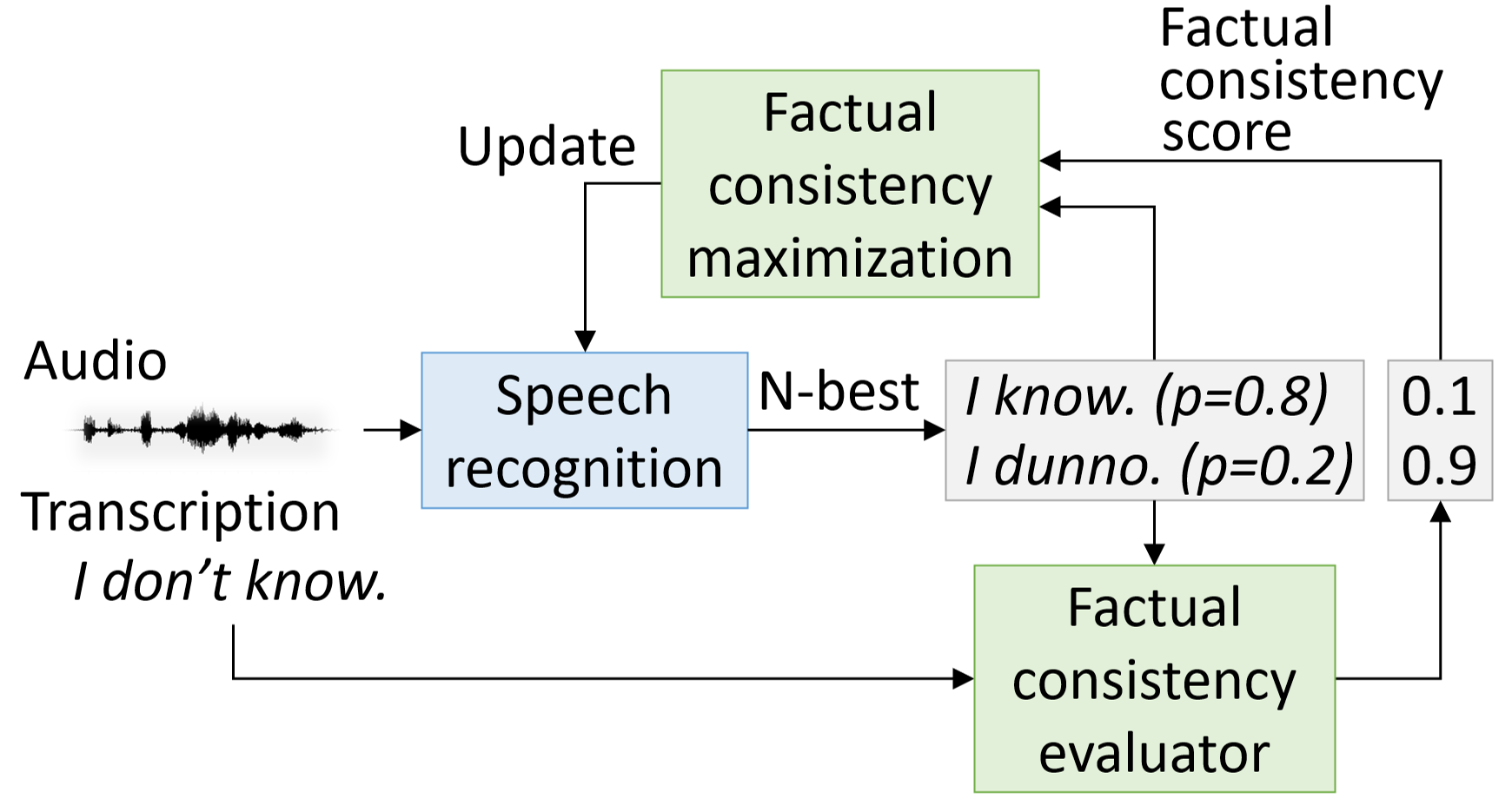}
   \vspace{-5mm}
      \caption{Factual consistency maximization of ASR. }
  \label{fig:overview}
    \vspace{-5mm}
\end{figure}

In this paper, we propose an ASR optimization scheme for factual consistency, with the aim of reducing hallucinations of the ASR model. Figure \ref{fig:overview} shows an overview of our proposed method along with illustrative transcriptions and scores. In this example, the ground-truth transcription of the input audio is ``{\it I don't know.}'' Meanwhile, the ASR system has produced ``{\it I know.}'' and ``{\it I dunno.}'' with posterior probabilities of 0.8 and 0.2, respectively. In terms of the WER ignoring the punctuations, 
the second hypothesis would be worse than the first one (67\% vs. 33\%).  However, 
the second hypothesis is factually or semantically more consistent with the ground-truth transcription. 
Our proposed optimization method encourages the ASR model to produce the second hypothesis over the first one by maximizing an expected factual consistency score.

Section \ref{sec:review} reviews related 
works. Section \ref{sec:proposed} describes the proposed factual consistency maximization framework. Section \ref{sec:experiment} reports  experimental results using the AMI meeting corpus \cite{carletta2005ami} and the VoxPopuli corpus \cite{wang2021voxpopuli}, where the effectiveness of our method is demonstrated for both utterance-wise evaluation and speech summarization. Section \ref{sec:conclusion} concludes the paper.

\section{Related works}
\label{sec:review}

\subsection{Quality measurement for generated texts}

A lot of effort has been made for measuring the quality of texts generated by ASR or NLP models. While the ASR models are typically evaluated with the WER or character error rate depending on the language, several task-specific metrics were also utilized, such as argument WER \cite{haghani2018audio} for spoken language understanding tasks. In NLP, various quality metrics were proposed for different tasks, including BLEU \cite{papineni2002bleu}, ROUGE \cite{lin2004rouge}, and METEOR \cite{banerjee2005meteor}. Recent work showed metrics based on pre-trained LMs are better correlated with human evaluation scores. BERTScore \cite{zhang2019bertscore}, MoverScore \cite{zhao2019moverscore}, and several other methods were proposed to estimate a general semantic similarity between two sentences based on pre-trained LMs.

{\it Factual consistency} between two texts \cite{kryscinski2020evaluating, wang2020asking, cao2020factual}, which is commonly used for evaluating natural language generation systems, 
is one key approach to taking account of hallucinations in the text quality assessment. 
In this work, we mainly use UniEval \cite{zhong2022towards} to compute the factual consistency score because of its superior correlation with human ratings. 
UniEval was trained by fine-tuning T5-large \cite{raffel2020exploring} based on artificially manipulated documents to provide multiple explainable dimensions of generated texts 
such as consistency
and fluency, resulting in multiple models with different flavors. 
UniEval-sum was specially trained for summarization to evaluate multiple
aspects related to summarization (e.g., consistency, fluency, coherence) while
UniEval-fact was trained only for factual consistency scoring.
Both models produce a factual consistency score
 between 0 (low consistency) and 1 (high consistency) given two texts.
In our work, we used UniEval-fact for ASR optimization and utterance-wise evaluation
while UniEval-sum for speech summarization evaluation.
We also used FactCC \cite{kryscinski2020evaluating}, which is another 
factual consistency evaluator based on BERT \cite{kenton2019bert},
in the utterance-wise evaluation
to further validate the effectiveness of the proposed method.

\subsection{ASR optimization}

An ASR model is usually trained with a cross-entropy (CE) loss between the model's output distribution and the ground-truth labels. 
Since the CE loss does not directly minimize the WER, alternative training schemes such as minimum WER training  \cite{graves2014towards,shannon2017optimizing,prabhavalkar2018minimum} are also widely adopted.
Our work can be considered as a variant of minimum WER training, where we optimize the ASR model based on the UniEval-based factual consistency score instead of the WER.

Joint optimization of ASR and NLP models, such as  end-to-end speech summarization \cite{sharma2022end,shon2022slue} or end-to-end spoken natural language understanding \cite{haghani2018audio,rao2020speech}, is also related to our work. 
Notably, Rao et al. \cite{rao2021mean} proposed to integrate evaluation metrics for intent classification and slot filling into the ASR training criterion. While Rao's work shares the same backbone in the optimization scheme with ours, we attempt to avoid using a task-specific metric and conduct evaluation with a task that the model is not directly trained for. 

\section{Factual consistency maximization training of ASR}
\label{sec:proposed}

\subsection{Training objective function}

Suppose we have
ASR training dataset $\mathcal{D}=\{X_r, Y_r\}_{r \in \Omega}$, 
where $X_r$ is the $r$-th training audio sample, $Y_r$ is the corresponding ground-truth transcription, and 
$\Omega$ is the sample index set. 
We propose to maximize the following training criterion: 
\begin{align}
&\mathcal{F}^{\scriptscriptstyle \mathrm{FC}}=\sum_{r \in \Omega} \bar{\mathcal{C}_r},
\end{align}
where
$\bar{\mathcal{C}_r}$ is the expected consistency score for the $r$-th training sample, which is estimated as follows:
\begin{align}
\bar{\mathcal{C}_r}=
    \sum_{Y\in \mathcal{B}(X_r)} \hat{P}(Y|X_r)\{|Y_r|\cdot \mathrm{Consistency}(Y; Y_r)\}.  \label{eq:ave}
\end{align}
Here, 
$\mathcal{B}()$ is a beam search function which returns the N-best hypotheses for $X_r$ based on the 
current ASR model parameters.
Function $\mathrm{Consistency}(Y;Y_r)$ calculates a consistency score of $Y$ measured against the ground-truth transcription $Y_r$, which is obtained by a factual consistency evaluator.
$|Y_r|$ is the number of words in $Y_r$. 
Finally, $\hat{P}(Y|X_r)$ is the normalized posterior probability of $Y$ that is defined as
\begin{align}
\hat{P}(Y|X_r)=\frac{P(Y|X_r)}
    {\sum_{Y'\in \mathcal{B}(X_r)} P(Y'|X_r)}, \label{eq:norm_prob}
\end{align}
where $P(Y|X_r)$ is the raw posterior probability of $Y$ given $X_r$.

\subsection{Training procedure}

Now, we describe the training procedure for attention encoder-decoder-based ASR models \cite{chorowski2014end}.
The training algorithms for other ASR model types, such as connectionist temporal classification \cite{graves2006connectionist} or recurrent neural network transducer \cite{graves2012sequence}, can be derived based on similar formulations.

With attention encoder-decoder-based ASR, the decoder module iteratively generates an output distribution over recognition tokens (i.e., characters, subwords, or words). Given the output $o_{n,i}$, where $n$ is the decoder iteration index and $i$ is the token index, the derivative of the objective function with respect to $\log(o_{n,i})$ is calculated for each $N$-best hypothesis $Y \in \mathcal{B}(X_r)$ if and only if the token index at the $n$-th position of $Y$ is equal to $i$.
\begin{align}
\frac{\partial \mathcal{F}^{\scriptscriptstyle \mathrm{FC}}}{\partial \log(o_{n,i})}=
\hat{P}(Y|X_r)\{
|Y_r|\cdot\mathrm{Consistency}(Y;Y_r)
-\bar{\mathcal{C}_r} 
\}. \label{eq:err} 
\end{align}
Otherwise, $\log(o_{n,i})$ is zero.

%
%
%
%
%

The overall training procedure is as follows. 
For each training sample $\{X_r, Y_r\}$, we first decode $X_r$ with beam search to generate the N-best hypotheses $Y$ based on the current model parameters. We then compute the normalized posterior probability of each hypothesis based on Eq. \eqref{eq:norm_prob}. Next, we compute the empirical expected consistency score $\bar{\mathcal{C}_r}$ per Eq. \eqref{eq:ave}. We then compute the training objective derivative with respect to $\log(o_{n,i})$ by using Eq. \eqref{eq:err}. 
Finally, we perform back-propagation to compute the gradient, and update the model parameter values based on the gradient ascent
algorithm. 

Note that 
maximizing the factual consistency does not necessarily improve the WER
as illustrated in Fig. \ref{fig:overview}.
Also, the factual consistency maximization tends to encourage
 a hypothesis with fewer output tokens to avoid hallucination errors.
Therefore, it is important to have some safeguard to
avoid severe deletion errors, 
such as interpolating a conventional CE loss.
In this work, we adopted a simple early-stopping approach 
where we applied the factual consistency maximization to a well-trained 
model based on the CE loss, and limited the number of updates 
to a relatively small number.

\section{Experiments}
\label{sec:experiment}

We evaluated the proposed ASR training framework in two scenarios. In the first experiment, both the training and evaluation of ASR models used short utterances (Section \ref{sec:ex-utt}) consistently. This experiment was carried out to validate the correctness and effectiveness of our formulation. In the second experiment, we conducted a speech summarization experiment (Section \ref{sec:ex-sum}). We applied speech summarization to the meeting transcriptions generated by the ASR models and measured the impacts of the different ASR models on the summarization quality. 
 This experiment was performed to investigate whether optimizing the ASR models for factual consistency helps improve a downstream task by using the speech summarization as an example. 
 Note that improving the summarization quality is not straightforward since 
 the ASR models are optimized at the utterance level despite the evaluation being performed with much longer segments.
 We also used a state-of-the-art large language model-based (LLM) summarization model to obtain relevant conclusions, unlike previous speech summarization works.

\subsection{Utterance-wise evaluation}
\label{sec:ex-utt}
\subsubsection{Experiment settings}

In this experiment, we evaluated ASR models by using 
both the utterance-level factual consistency score and WER to examine 
the impact of the proposed training method. 
We used the AMI meeting corpus \cite{carletta2005ami} and the VoxPopuli corpus \cite{wang2021voxpopuli}.

The AMI meeting corpus contains approximately 100 hours of meeting recordings captured by both independent headset microphones (IHM) and multiple distant microphones. In our experiment, we used the IHM audio with ground-truth utterance boundaries. We followed the Kaldi recipe \cite{povey2011kaldi} for partitioning the data into training, development, and evaluation sets, resulting in 80.2 hours, 9.7 hours, and 9.1 hours of recordings for these splits, respectively. Unlike a conventional ASR experiment using case-normalized and non-punctuated transcriptions for training and testing, we used the case-sensitive and punctuated-transcriptions based on the official transcription\footnote{We used the words and punctuations of the official annotation without modifications. The only exception was that we removed underscores from a word (e.g., X\_M\_L\_ was converted to XML).} throughout our experiments. We chose to do so because the factual consistency evaluator (in our case, UniEval \cite{zhong2022towards}) is case- and punctuation-sensitive. Only for the WER scoring, we applied the text normalizer provided by \cite{radford2022robust}.

The VoxPopuli corpus consists of 100K hours of unlabelled speech data in 23 languages, as well as 1.8K hours of transcribed speech in 16 languages. In our experiment, we used the transcribed English subset. 
The official dataset is split into training, development, and testing sets, which contains 522.6 hours, 5.0 hours, and 4.9 hours of speech, respectively. As with the AMI meeting corpus, we used the case-sensitive, punctuated transcriptions. Note that we excluded roughly 3\% of the development and testing sets (58 out of 1753 sentences in the development set and 52 out of 1842 sentences in the testing set) that did not have case-sensitive, punctuated transcriptions.

We used the Whisper Base model \cite{radford2022robust} as our initial ASR model. In the experiment with the AMI meeting corpus, we first fine-tuned the Whisper Base on the AMI training set with the CE loss. We conducted 2,500 training iterations with 8 GPUs, each consuming mini-batches of 15,000 frames.
We used a linear decay learning rate schedule with an initial learning rate of 1e-5. After the CE-loss-based fine-tuning, we further updated the model based on the proposed factual consistency maximization by using the same AMI training data. We used UniEval-fact
to compute the consistency score and used an N-best size of 4 to compute the gradient. We conducted 2,500 training iterations with 8 GPUs, each consuming 1 sample for one training iteration. A linear decay learning rate schedule was used, starting with a learning rate of 1e-6. The VoxPopuli experiment was conducted with almost the same setting but using 6,250 training iterations with 8 GPUs (i.e., a total of 50,000-minibatch consumption) for both the CE loss-based training and factual consistency maximization.

In the evaluation, we performed beam search decoding with a beam size of 4. We set the language tag to English to enforce the model to output only English transcriptions \cite{radford2022robust}. We evaluated the WER by applying the text normalizer provided in \cite{radford2022robust}. We also computed the factual consistency score of the ASR hypothesis given the ground-truth transcription using two different factual consistency evaluators:  UniEval-fact \cite{zhong2022towards} and FactCC \cite{kryscinski2020evaluating}. For UniEval-fact, we report the average factual consistency score over the test utterances. For FactCC, we report the ratio of the hypotheses that were judged to be consistent with the ground-truth transcriptions as FactCC makes a binary decision as to the factual consistency. For both consistency metrics, a higher score means the ASR hypothesis is more factually consistent with the ground-truth transcription.


\begin{table}[t]
\ra{0.9}
\tabcolsep = 0.7mm
  \caption{WER (\%), UniEval-fact consistency score (UE) and FactCC consistency score (FCC) for AMI-IHM development and evaluation sets. 
 FCM: factual consistency maximization.}
  \label{tab:ami-ihm}
  \vspace{-3mm}
  \centering
{   \footnotesize
\begin{tabular}{@{}lccccccc@{}}
    \toprule
\multicolumn{1}{c}{ASR model} & \multicolumn{3}{c}{AMI-IHM dev} && \multicolumn{3}{c}{AMI-IHM eval} \\ \cmidrule{2-4} \cmidrule{6-8}
             & WER {\scriptsize ($\downarrow$)} & UE {\scriptsize ($\uparrow$)} & FCC {\scriptsize ($\uparrow$)} && WER {\scriptsize ($\downarrow$)}& UE {\scriptsize ($\uparrow$)} & FCC {\scriptsize ($\uparrow$)} \\ \midrule
Whisper Base                                  & 19.7 & 0.727 & 0.925  && 20.3  & 0.719 & 0.913 \\ 
\hspace{3mm}$\hookrightarrow$ CE-loss    & 11.5 & 0.785 & 0.930 && 12.6 & 0.787 & 0.932\\ 
\hspace{6mm}$\hookrightarrow$ FCM  & {\bf 11.4} & {\bf 0.799} & {\bf 0.942} && {\bf 12.5}  & {\bf 0.801} & {\bf 0.940} \\ \bottomrule
  \end{tabular}
  }
  \vspace{-2mm}
\end{table}

\begin{table}[t]
\ra{0.9}
\tabcolsep = 0.7mm
  \caption{WER (\%), UniEval-fact consistency score (UE) and FactCC consistency score (FCC) for VoxPopuli development and test sets. 
 FCM: factual consistency maximization.}
  \label{tab:voxpopuli}
  \vspace{-3mm}
  \centering
{   \footnotesize
\begin{tabular}{@{}lccccccc@{}}
    \toprule
\multicolumn{1}{c}{ASR model} & \multicolumn{3}{c}{VoxPopuli dev$^\star$} && \multicolumn{3}{c}{VoxPopuli test$^\star$} \\ \cmidrule{2-4} \cmidrule{6-8}
             & WER {\scriptsize ($\downarrow$)} & UE {\scriptsize ($\uparrow$)} & FCC {\scriptsize ($\uparrow$)}
 && WER {\scriptsize ($\downarrow$)}& UE {\scriptsize ($\uparrow$)} & FCC {\scriptsize ($\uparrow$)}\\ \midrule
Whisper Base                                  & 9.4 & 0.755 & 0.896 && 9.4  & 0.747 & 0.887 \\ 
\hspace{3mm}$\hookrightarrow$ CE-loss     & {\bf 8.2} & 0.766 & 0.901 && {\bf 8.3}  & 0.757 & 0.886 \\ 
\hspace{6mm}$\hookrightarrow$ FCM & 8.4 & {\bf 0.795} & {\bf 0.920} && 8.5  & {\bf 0.791} & {\bf 0.911} \\ \bottomrule
  \end{tabular}
  }
  \\{\scriptsize $^{\star}$ Utterances without case and punctuated transcriptions were excluded.}
  \vspace{-5mm}
\end{table}

\subsubsection{Results}

The results for AMI-IHM and VoxPopuli are shown in Tables \ref{tab:ami-ihm} and \ref{tab:voxpopuli}, respectively. We can see that the CE-loss-based fine-tuning achieved significant gains in both WER and consistency score for most evaluation sets except for the FCC score on VoxPopuli test set. Furthermore, we can see that the proposed factual consistency maximization further improved all the consistency metrics while keeping the WER almost intact. 
For VoxPopuli, as one may expect, marginal WER degradation was observed because our training objective was not aimed at minimizing the label prediction errors as illustrated in Fig. \ref{fig:overview}.

\subsection{Speech summarization evaluation}
\label{sec:ex-sum}
\subsubsection{Experiment settings}

We also evaluated the proposed ASR training method by performing automatic summarization using the ASR-generated transcriptions. In this experiment, we used the AMI meeting corpus. 

We opted to create our own experiment setup due to the challenges that we faced when we attempted to use existing speech summarization corpora. 
We initially tried to use the ROUGE score \cite{lin2004rouge} based on the human-crafted abstractive summaries provided in the AMI official annotations. However, a preliminary experiment revealed that the number of sessions in the AMI development and evaluation sets was too small (18 and 16 sessions, respectively) 
to discuss the statistical differences in the summarization quality between different systems. 
We also considered other speech summarization datasets, namely the How2 dataset \cite{sanabria18how2}
and the SLUE-TED corpus \cite{shon2022slue}.
However, the How2 dataset provides only feature files that are not compatible with our ASR models,
and the SLUE-TED corpus has yet to be released at the time of writing.
In addition to the lack of the evaluation data, recent studies also reported that the correlation between the ROUGE and human scores is much weaker than LM-based metrics (e.g., \cite{zhong2022towards}).

With these considerations, we designed our speech summarization experiment as follows.\vspace{-1mm}\\
\vspace{-0mm}\hspace{0mm}\rule{0.47\textwidth}{0.4pt}\vspace{-1mm}
\begin{enumerate}
    \item Split the AMI development and evaluation set recordings into non-overlapping chunks of 60 seconds.
    \item Conduct the following steps for each 60-s chunk.
\begin{enumerate}
    \item Apply ASR. We used the IHM audio with ground-truth segmentation.
    \item Convert the ASR hypotheses to a speaker-attributed transcription format with a template of ``Speaker [mic-index]: [ASR-hypothesis]\textbackslash n''. 
    We used the headset microphone index as the speaker index and sorted the hypotheses based on the start time of each utterance.
For example, three 
ASR hypotheses consisting of 
``Hello. (mic-index=1, start=1s)'', 
``Hi, how are you? (mic-index=2, start=3s)'', and ``I'm fine. (mic-indexl=1, start=5s)''
would be converted to 
        ``Speaker 1: Hello.\textbackslash nSpeaker 2: Hi, how are you?\textbackslash nSpeaker 1: I'm fine.\textbackslash n''. 
    \item Generate a summary text for the speaker-attributed transcription obtained in the previous step. 
We used InstructGPT (text-davinci-002) \cite{ouyang2022training} to 
generate the summary by prompting the model as  
``[speaker-attributed transcription]\textbackslash nSummarize 
the conversation above.\textbackslash n''. We employed the least randomness configuration by setting temperature, top\_p, and max\_tokens at 0.0, 1.0, and 200, respectively.
    \item Compute the factual consistency score of the generated summary by using the ground-truth speaker-attributed transcription as the reference, which was formatted in the same way as 2 (b).  We used 
    UniEval-sum \cite{zhong2022towards} for the consistency scoring.
\end{enumerate}
   \item Compute the average of the factual consistency scores over all the 60-s chunks.
\end{enumerate}
\vspace{-3mm}\rule{0.47\textwidth}{0.4pt}

As a result, we obtained 574 and 542 speech summarization testing materials (i.e., 60-s chunks) for the development and evaluation sets, respectively, which enabled us to draw statistically relevant conclusions regarding the summarization quality. Note that UniEval-sum can also produce other metrics, such as the fluency and coherence of the generated summaries. As these metrics are irrelevant to the ASR errors (that is, the fluency and coherence scores are determined primarily by the summarization model quality rather than the quality of the input text produced by the ASR), we only report the UniEval-sum factual consistency result, which would best reflect the impact of the ASR performance on the summarization.

%

\subsubsection{Results}

Table \ref{tab:ami-sum} shows the speech summarization experiment result. The ASR models trained in the utterance-wise evaluation experiment (Section \ref{sec:ex-utt}) were used. We also measured the consistency score for the summaries generated by using the ground-truth transcriptions as input. This can be considered as 
the upper bound of the consistency score
that can be achieved with the employed summarization model. First, we can observe that the summarization consistency score obtained with Whisper Base was significantly lower than that of the ground-truth transcription. This shows the negative impact of the ASR errors on the summarization quality. 
It can also be seen that the CE fine-tuning slightly improved the consistency score. The consistency score for the development set was only improved by 0.1\% despite the significant WER improvement from 19.7\% to 11.5\% (Table \ref{tab:ami-ihm}). 
On the other hand, the proposed factual consistency maximization further improved the consistency scores for both the development and evaluation sets, coming close to the ground-truth-based upper bound scores. 

\begin{table}[t]
\ra{0.9}
\tabcolsep = 3.0mm
  \caption{Summarization consistency score based on UniEval-sum for AMI development and evaluation sets. FC: factual consistency.}
  \label{tab:ami-sum}
  \vspace{-3mm}
  \centering
{   \footnotesize
\begin{tabular}{@{}lccccc@{}}
    \toprule
\multicolumn{1}{c}{ASR model} & \multicolumn{2}{c}{Summarization consistency {\scriptsize ($\uparrow$)}}  \\ \cmidrule{2-3} 
     & AMI dev & AMI eval  \\ \midrule
Whisper Base                                  & 0.697 & 0.739 \\ 
\hspace{2mm}$\hookrightarrow$ CE-loss fine-tuning    & 0.698 & 0.745 \\ 
\hspace{4mm}$\hookrightarrow$ FC maximization  & {\bf 0.706} & {\bf 0.748} \\  \hdashline[1pt/2pt]\hdashline[0pt/1pt] 
Ground-truth transcription  & 0.709 & 0.753 \\ \bottomrule

  \end{tabular}
  }
  \vspace{-5mm}
\end{table}

We further conducted paired t-tests for the results of Table \ref{tab:ami-sum}. 
The consistency score difference between the summary texts obtained with Whisper Base (0.739) and those with the CE fine-tuned model (0.745) was judged to be not significant with 95\% confidence. On the other hand, the difference between the summarization with Whisper Base (0.739) and that with the proposed factual consistency-maximized model (0.748) was judged to be significant with 95\% confidence.
Finally, the difference between the summarization with the CE fine-tuned model and that with the factual consistency-maximized model was also judged to be significant with 95\% confidence for the development set (0.698 vs 0.706) while it was not the case for the evaluation set (0.745 vs 0.748), which we would attribute to the yet limited number of evaluation samples. Overall, we observed a clear trend showing the effectiveness of the proposed factual consistency maximization method even when using the powerful LLM for the speech summarization.

\section{Conclusions}
\label{sec:conclusion}

In this paper, we proposed a novel ASR training framework, which optimizes the ASR model for maximizing the factual consistency between an ASR hypothesis and a ground-truth transcription. In our experiments using the AMI meeting corpus and the VoxPopuli corpus, we showed that the ASR model optimized by the proposed framework produced hypotheses which had significantly higher consistency scores with the ground-truth transcriptions while keeping the WER almost intact. 
We also showed that using the ASR models trained with the proposed framework improved the speech summarization quality for meeting conversations, demonstrating the usefulness of the proposed method in a challenging downstream NLP task.

\bibliographystyle{IEEEtran}
\bibliography{mybib}

\begin{thebibliography}{10}
\providecommand{\url}[1]{#1}
\csname url@samestyle\endcsname
\providecommand{\newblock}{\relax}
\providecommand{\bibinfo}[2]{#2}
\providecommand{\BIBentrySTDinterwordspacing}{\spaceskip=0pt\relax}
\providecommand{\BIBentryALTinterwordstretchfactor}{4}
\providecommand{\BIBentryALTinterwordspacing}{\spaceskip=\fontdimen2\font plus
\BIBentryALTinterwordstretchfactor\fontdimen3\font minus
  \fontdimen4\font\relax}
\providecommand{\BIBforeignlanguage}[2]{{%
\expandafter\ifx\csname l@#1\endcsname\relax
\typeout{** WARNING: IEEEtran.bst: No hyphenation pattern has been}%
\typeout{** loaded for the language `#1'. Using the pattern for}%
\typeout{** the default language instead.}%
\else
\language=\csname l@#1\endcsname
\fi
#2}}
\providecommand{\BIBdecl}{\relax}
\BIBdecl

\bibitem{vaswani2017attention}
A.~Vaswani, N.~Shazeer, N.~Parmar \emph{et~al.}, ``Attention is all you need,''
  in \emph{Proc. NIPS}, 2017, pp. 6000--6010.

\bibitem{gulati2020conformer}
A.~Gulati, J.~Qin, C.-C. Chiu \emph{et~al.}, ``Conformer: Convolution-augmented
  {Transformer} for speech recognition,'' \emph{Proc. Interspeech}, pp.
  5036--5040, 2020.

\bibitem{graves2014towards}
A.~Graves and N.~Jaitly, ``Towards end-to-end speech recognition with recurrent
  neural networks,'' in \emph{Proc. ICML}.\hskip 1em plus 0.5em minus
  0.4em\relax PMLR, 2014, pp. 1764--1772.

\bibitem{povey2016purely}
D.~Povey, V.~Peddinti, D.~Galvez \emph{et~al.}, ``Purely sequence-trained
  neural networks for {ASR} based on lattice-free {MMI},'' in \emph{Proc.
  Interspeech}, 2016, pp. 2751--2755.

\bibitem{shannon2017optimizing}
M.~Shannon, ``Optimizing expected word error rate via sampling for speech
  recognition,'' \emph{Proc. Interspeech}, pp. 3537--3541, 2017.

\bibitem{prabhavalkar2018minimum}
R.~Prabhavalkar, T.~N. Sainath, Y.~Wu \emph{et~al.}, ``Minimum word error rate
  training for attention-based sequence-to-sequence models,'' in \emph{Proc.
  ICASSP}, 2018, pp. 4839--4843.

\bibitem{parthasarathi2019lessons}
S.~H.~K. Parthasarathi and N.~Strom, ``Lessons from building acoustic models
  with a million hours of speech,'' in \emph{Proc. ICASSP}, 2019, pp.
  6670--6674.

\bibitem{kanda2021large}
N.~Kanda, G.~Ye, Y.~Wu \emph{et~al.}, ``Large-scale pre-training of end-to-end
  multi-talker {ASR} for meeting transcription with single distant
  microphone,'' in \emph{Proc. Interspeech}, 2021, pp. 3430--3434.

\bibitem{zhang2022bigssl}
Y.~Zhang, D.~S. Park, W.~Han \emph{et~al.}, ``{BigSSL}: Exploring the frontier
  of large-scale semi-supervised learning for automatic speech recognition,''
  \emph{IEEE Journal of Selected Topics in Signal Processing}, 2022.

\bibitem{radford2022robust}
A.~Radford, J.~W. Kim, T.~Xu \emph{et~al.}, ``Robust speech recognition via
  large-scale weak supervision,'' \emph{arXiv preprint arXiv:2212.04356}, 2022.

\bibitem{panayotov2015librispeech}
V.~Panayotov, G.~Chen, D.~Povey, and S.~Khudanpur, ``Librispeech: an {ASR}
  corpus based on public domain audio books,'' in \emph{Proc. ICASSP}, 2015,
  pp. 5206--5210.

\bibitem{amodei2016deep}
D.~Amodei, S.~Ananthanarayanan, R.~Anubhai \emph{et~al.}, ``{Deep Speech} 2:
  End-to-end speech recognition in {English} and {Mandarin},'' in \emph{Proc.
  ICML}, 2016, pp. 173--182.

\bibitem{godfrey1992switchboard}
J.~J. Godfrey, E.~C. Holliman, and J.~McDaniel, ``{SWITCHBOARD}: Telephone
  speech corpus for research and development,'' in \emph{Proc. ICASSP}, vol.~1,
  1992, pp. 517--520.

\bibitem{xiong2016achieving}
W.~Xiong, J.~Droppo, X.~Huang \emph{et~al.}, ``Achieving human parity in
  conversational speech recognition,'' \emph{arXiv preprint arXiv:1610.05256},
  2016.

\bibitem{graves2006connectionist}
A.~Graves, S.~Fern{\'a}ndez, F.~Gomez, and J.~Schmidhuber, ``Connectionist
  temporal classification: labelling unsegmented sequence data with recurrent
  neural networks,'' in \emph{Proc. ICML}, 2006, pp. 369--376.

\bibitem{graves2012sequence}
A.~Graves, ``Sequence transduction with recurrent neural networks,''
  \emph{arXiv preprint arXiv:1211.3711}, 2012.

\bibitem{chorowski2014end}
J.~Chorowski, D.~Bahdanau, K.~Cho, and Y.~Bengio, ``End-to-end continuous
  speech recognition using attention-based recurrent {NN}: First results,'' in
  \emph{NIPS Workshop on Deep Learning}, 2014.

\bibitem{seide2011conversational}
F.~Seide, G.~Li, and D.~Yu, ``Conversational speech transcription using
  context-dependent deep neural networks,'' in \emph{Proc. Interspeech}, 2011,
  pp. 437--440.

\bibitem{hakkani2006beyond}
D.~Hakkani-T{\"u}r, F.~B{\'e}chet, G.~Riccardi, and G.~Tur, ``Beyond {ASR}
  1-best: Using word confusion networks in spoken language understanding,''
  \emph{Computer Speech \& Language}, vol.~20, no.~4, pp. 495--514, 2006.

\bibitem{morbini2012reranking}
F.~Morbini, K.~Audhkhasi, R.~Artstein \emph{et~al.}, ``A reranking approach for
  recognition and classification of speech input in conversational dialogue
  systems,'' in \emph{Proc. SLT}, 2012, pp. 49--54.

\bibitem{brown2020language}
T.~Brown, B.~Mann, N.~Ryder \emph{et~al.}, ``Language models are few-shot
  learners,'' \emph{Advances in neural information processing systems},
  vol.~33, pp. 1877--1901, 2020.

\bibitem{haghani2018audio}
P.~Haghani, A.~Narayanan, M.~Bacchiani \emph{et~al.}, ``From audio to
  semantics: Approaches to end-to-end spoken language understanding,'' in
  \emph{Proc. SLT}, 2018, pp. 720--726.

\bibitem{rao2020speech}
M.~Rao, A.~Raju, P.~Dheram \emph{et~al.}, ``Speech to semantics: Improve {ASR}
  and {NLU} jointly via all-neural interfaces,'' in \emph{Proc. Interspeech},
  2020, pp. 876--880.

\bibitem{sharma2022end}
R.~Sharma, S.~Palaskar, A.~W. Black, and F.~Metze, ``End-to-end speech
  summarization using restricted self-attention,'' in \emph{Proc. ICASSP},
  2022, pp. 8072--8076.

\bibitem{shon2022slue}
S.~Shon, S.~Arora, C.-J. Lin \emph{et~al.}, ``{SLUE} phase-2: A benchmark suite
  of diverse spoken language understanding tasks,'' \emph{arXiv preprint
  arXiv:2212.10525}, 2022.

\bibitem{rao2021mean}
M.~Rao, P.~Dheram, G.~Tiwari \emph{et~al.}, ``Do as i mean, not as i say:
  Sequence loss training for spoken language understanding,'' in \emph{Proc.
  ICASSP}, 2021, pp. 7473--7477.

\bibitem{carletta2005ami}
J.~Carletta, S.~Ashby, S.~Bourban \emph{et~al.}, ``The {AMI} meeting corpus: A
  pre-announcement,'' in \emph{International workshop on machine learning for
  multimodal interaction}, 2006, pp. 28--39.

\bibitem{wang2021voxpopuli}
C.~Wang, M.~Riviere, A.~Lee \emph{et~al.}, ``{VoxPopuli}: A large-scale
  multilingual speech corpus for representation learning, semi-supervised
  learning and interpretation,'' in \emph{Proc. ACL-IJCNLP}, 2021, pp.
  993--1003.

\bibitem{papineni2002bleu}
K.~Papineni, S.~Roukos, T.~Ward, and W.-J. Zhu, ``{BLEU}: a method for
  automatic evaluation of machine translation,'' in \emph{Proc. ACL}, 2002, pp.
  311--318.

\bibitem{lin2004rouge}
C.-Y. Lin, ``{ROUGE}: A package for automatic evaluation of summaries,'' in
  \emph{Text Summarization Branches Out}, 2004, pp. 74--81.

\bibitem{banerjee2005meteor}
S.~Banerjee and A.~Lavie, ``{METEOR}: An automatic metric for {MT} evaluation
  with improved correlation with human judgments,'' in \emph{Proc. of the ACL
  workshop on intrinsic and extrinsic evaluation measures for machine
  translation and/or summarization}, 2005, pp. 65--72.

\bibitem{zhang2019bertscore}
T.~Zhang, V.~Kishore, F.~Wu \emph{et~al.}, ``{BERTScore}: Evaluating text
  generation with {BERT},'' \emph{arXiv preprint arXiv:1904.09675}, 2019.

\bibitem{zhao2019moverscore}
W.~Zhao, M.~Peyrard, F.~Liu \emph{et~al.}, ``{MoverScore}: Text generation
  evaluating with contextualized embeddings and earth mover distance,'' in
  \emph{Proc. EMNLP-IJCNLP}, 2019, pp. 563--578.

\bibitem{kryscinski2020evaluating}
W.~Kry{\'s}ci{\'n}ski, B.~McCann, C.~Xiong, and R.~Socher, ``Evaluating the
  factual consistency of abstractive text summarization,'' in \emph{Proc.
  EMNLP}, 2020, pp. 9332--9346.

\bibitem{wang2020asking}
A.~Wang, K.~Cho, and M.~Lewis, ``Asking and answering questions to evaluate the
  factual consistency of summaries,'' \emph{arXiv preprint arXiv:2004.04228},
  2020.

\bibitem{cao2020factual}
M.~Cao, Y.~Dong, J.~Wu, and J.~C.~K. Cheung, ``Factual error correction for
  abstractive summarization models,'' in \emph{Proc. EMNLP}, 2020, pp.
  6251--6258.

\bibitem{zhong2022towards}
M.~Zhong, Y.~Liu, D.~Yin \emph{et~al.}, ``Towards a unified multi-dimensional
  evaluator for text generation,'' in \emph{Proc. EMNLP}, 2022, pp. 2023--2038.

\bibitem{raffel2020exploring}
C.~Raffel, N.~Shazeer, A.~Roberts \emph{et~al.}, ``Exploring the limits of
  transfer learning with a unified text-to-text transformer,'' \emph{The
  Journal of Machine Learning Research}, vol.~21, no.~1, pp. 5485--5551, 2020.

\bibitem{kenton2019bert}
J.~Devlin, M.-W. Chang, K.~Lee, and K.~Toutanova, ``{BERT}: Pre-training of
  deep bidirectional transformers for language understanding,'' in \emph{Proc.
  NAACL-HLT}, 2019, pp. 4171--4186.

\bibitem{povey2011kaldi}
D.~Povey, A.~Ghoshal, G.~Boulianne \emph{et~al.}, ``The {Kaldi} speech
  recognition toolkit,'' in \emph{Proc. ASRU}, 2011.

\bibitem{sanabria18how2}
R.~Sanabria, O.~Caglayan, S.~Palaskar \emph{et~al.}, ``{How2:} a large-scale
  dataset for multimodal language understanding,'' in \emph{Proceedings of the
  Workshop on Visually Grounded Interaction and Language}.\hskip 1em plus 0.5em
  minus 0.4em\relax NeurIPS, 2018.

\bibitem{ouyang2022training}
L.~Ouyang, J.~Wu, X.~Jiang \emph{et~al.}, ``Training language models to follow
  instructions with human feedback,'' \emph{arXiv preprint arXiv:2203.02155},
  2022.

\end{thebibliography}

\end{document}